\documentclass[journal]{IEEEtran}
\usepackage{ifpdf}

\usepackage{cite}

\ifCLASSINFOpdf
  \usepackage[pdftex]{graphicx}
\else
  \usepackage[dvips]{graphicx}
\fi
\usepackage{amsmath, amssymb}
\usepackage{algorithmic, algorithm}

\usepackage{array}

\ifCLASSOPTIONcompsoc
 \usepackage[caption=false,font=normalsize,labelfont=sf,textfont=sf]{subfig}
\else
 \usepackage[caption=false,font=footnotesize]{subfig}
\fi

\usepackage{stfloats}
\ifCLASSOPTIONcaptionsoff
 \usepackage[nomarkers]{endfloat}
\let\MYoriglatexcaption\caption
\renewcommand{\caption}[2][\relax]{\MYoriglatexcaption[#2]{#2}}
\fi
\usepackage{url}

\usepackage[colorlinks,linkcolor=red,anchorcolor=green,citecolor=blue]{hyperref}

\usepackage{multirow}
\usepackage{booktabs}
\usepackage{arydshln}
\usepackage[table,xcdraw]{xcolor}
\usepackage{longtable}
\usepackage{subfig}
\usepackage{subfloat}
\usepackage{booktabs}
\usepackage{threeparttable}
\usepackage{lscape}
\usepackage{threeparttable}
\usepackage{verbatim}
\usepackage{marvosym}
\usepackage{ulem}
\usepackage{fancyhdr}
\usepackage{soul}
\usepackage{pifont}
\usepackage{fontawesome}

\usepackage{graphicx} 
\newcommand{\instbit}[1]{\mbox{\scriptsize #1}}
\newcommand{\instbitrange}[2]{~\instbit{#1} \hfill \instbit{#2}~}
\usepackage{siunitx}

\usepackage{tabularray} 

\newcommand{\rOneLQ}[1]{\textcolor{black}{#1}}
\newcommand{\rOneFC}[1]{\textcolor{black}{#1}}
\newcommand{\rTwoLQ}[1]{\textcolor{black}{#1}}
\newcommand{\rThreeLQ}[1]{\textcolor{black}{#1}}
\newcommand{\rTwoFC}[1]{\textcolor{black}{#1}}
\newcommand{\rRevLQ}[1]{\textcolor{black}{#1}}

\begin{document}

%

\title{\huge Enable Lightweight and Precision-Scalable Posit/IEEE-754 Arithmetic in RISC-V Cores for Transprecision Computing}

%



\author{Qiong~Li$^\dagger$, Chao~Fang$^\dagger$,~\IEEEmembership{Graduate~Student~Member,~IEEE}, Longwei~Huang, \\
        Jun~Lin,~\IEEEmembership{Senior~Member,~IEEE}, 
        and~Zhongfeng~Wang,~\IEEEmembership{Fellow,~IEEE}
\thanks{$^\dagger$Equal contribution. This work was supported in part by the National Key R\&D Program of China under Grant 2022YFB4400600, in part by the National Natural Science Foundation of China under Grant 62174084, and in part by the Postgraduate Research \& Practice Innovation Program of Jiangsu Province under Grant SJCX23\_0016. \textit{(Corresponding author: Zhongfeng Wang.)}}

\thanks{Q. Li, C. Fang, L. Huang and J. Lin are with the School of Electronic Science and Engineering, Nanjing University (e-mail: qiongli@smail.nju.edu.cn; fantasysee@smail.nju.edu.cn; lwhuang@smail.nju.edu.cn; jlin@nju.edu.cn).}
\thanks{Z. Wang is with the School of Electronic Science and Engineering, Nanjing University, and the School of Integrated Circuits, Sun Yat-sen University (email: zfwang@nju.edu.cn).}
\vspace{-0.01cm}}

%
%

\markboth{Journal of \LaTeX\ Class Files,~Vol.~14, No.~8, August~2015}%
{Shell \MakeLowercase{\textit{et al.}}: Bare Demo of IEEEtran.cls for IEEE Journals}
%



\maketitle

\begin{abstract}
\rRevLQ{While posit format offers superior dynamic range and accuracy for transprecision computing, its adoption in RISC-V processors is hindered by the lack of a unified solution for lightweight, precision-scalable, and IEEE-754 arithmetic compatible hardware implementation.
To address these challenges, we enhance RISC-V processors by 1) integrating dedicated posit codecs into the original FPU for lightweight implementation, 2) incorporating multi/mixed-precision support with dynamic exponent size for precision-scalability, and 3) reusing and customizing ISA extensions for IEEE-754 compatible posit operations.
Our comprehensive evaluation spans the modified FPU, RISC-V core, and SoC levels. 
It demonstrates that our implementation achieves 47.9\% LUTs and 57.4\% FFs reduction compared to state-of-the-art posit-enabled RISC-V processors, while achieving up to 2.54$\times$  throughput improvement in various GEMM kernels.}
\end{abstract}

\begin{IEEEkeywords}
Posit, RISC-V, precision-scalable, transprecision computing, IEEE-754, custom instructions.
\end{IEEEkeywords}

%
\IEEEpeerreviewmaketitle

\vspace{-0.5em}
\section{Introduction} \label{sec:intro}
\IEEEPARstart{T}{ransprecision} \rOneFC{computing~\cite{malossi2018transprecision} leverages variable precisions for diverse application needs~\cite{huang2024precision}, targeting reduced memory footprint and \rThreeLQ{computational complexity} without compromising accuracy.}
\rThreeLQ{This approach has received increasing attention in modern computing systems, as high-precision computation everywhere incurs significant storage overhead and memory access energy costs.}
\rOneLQ{However, traditional low-precision floating-point (FP) formats are susceptible to accuracy degradation \cite{gustafson2017beating}.}
\rRevLQ{Alternatively, posit format \cite{gustafson2017beating} is a promising replacement due to its superior dynamic range and numerical accuracy, improving computational accuracy than the IEEE-754 counterpart~\cite{lu2020evaluations}.}

\rRevLQ{Nevertheless, existing studies~\cite{mallasen2022percival, sharma2023clarinet, cococcioni2021lightweight, tiwari2021peri, arunkumar2020perc, oh2023rf2p, ciocirlan2021accuracy, rossi2023fppu} integrating posit arithmetic into RISC-V cores face several challenges in hardware efficiency, cross-precision capability, and IEEE-754 compatibility, particularly for transprecision computing applications.}
\rTwoFC{\rRevLQ{Regarding hardware efficiency}, some works~\cite{mallasen2022percival, sharma2023clarinet} incorporate a complete posit arithmetic unit (PAU) within the execution pipeline stage parallel to the original floating-point unit (FPU), resulting in significant hardware overhead.}
\rTwoFC{\rRevLQ{In terms of cross-precision capability}, \rThreeLQ{prior works~\cite{mallasen2022percival, rossi2023fppu, ciocirlan2021accuracy, oh2023rf2p} generally} lack support for multiple and mixed-precision \cite{micikevicius2017mixed} posit operations, limiting their effectiveness across various transprecision scenarios.}
\rTwoFC{Additionally, these works fail to support dynamic exponent size ($es$), a crucial feature for optimizing the accuracy-range trade-off benefits of posit.}
\rTwoFC{\rRevLQ{Concerning IEEE-754 compatibility}, previous approaches~\cite{rossi2023fppu, tiwari2021peri, oh2023rf2p, arunkumar2020perc, ciocirlan2021accuracy} support posit arithmetic by replacing the FPU with a PAU, sacrificing compatibility with standard FP formats essential for general-purpose transprecision computing.}
\rTwoFC{Moreover, to ensure IEEE-754 compatibility while incorporating posit arithmetic, there is a critical need for carefully designed ISA extensions for posit operations.}
\rThreeLQ{However, \cite{mallasen2022percival} fails to address the necessary conversion between posit and FP formats within its proposed instruction set, 
and \cite{arunkumar2020perc} lacks complete instruction support.}
\rOneFC{To tackle the above challenges}, we implement lightweight and precision-scalable posit/IEEE-754 arithmetic in RISC-V cores,
which enables efficient transprecision computing.
\rThreeLQ{Specifically, we modify the built-in FPU by integrating dedicated codecs to facilitate posit support, while maintaining IEEE-754 compatibility by skipping these codecs.}
\rThreeLQ{This strategy minimizes hardware overhead and ensures our design's lightweight nature.}
\rThreeLQ{Moreover, we enhance the microarchitecture to support multi/mixed-precision operations and further implement dynamic $es$ in hardware for posits, 
empowering our design with precision scalability to deal with various transprecision scenarios.}
\rRevLQ{We leverage existing instructions to perform standalone posit/IEEE-754 operations, 
and handle their compatibility by customizing ISA extensions that implement conversion between different data types and precisions.}
Finally, we incorporate the modified core into PULP SoC \cite{pullini2019mr} and conduct a \rThreeLQ{comprehensive} evaluation of our modifications at the FPU, RISC-V core, and SoC levels.

\rRevLQ{To summarize, our contributions are as follows.}

\begin{enumerate}
    \item \rRevLQ{Lightweight posit implementation on RISC-V cores by integrating dedicated codecs into the original FPU.}
    \rRevLQ{This significantly reduces 47.9\% LUTs and 57.4\% FFs compared to prior work \cite{mallasen2022percival}.}
    \item \rRevLQ{Efficient precision-scalable posit computing by supporting multi/mixed-precision operations and enabling dynamic exponent size through microarchitectural enhancements.}
    \rRevLQ{Our design outperforms existing works \cite{tiwari2021peri, arunkumar2020perc, oh2023rf2p, mallasen2022percival, ciocirlan2021accuracy, rossi2023fppu, sharma2023clarinet, cococcioni2021lightweight} with superior transprecision capabilities.}
    \item \rRevLQ{IEEE-754 compatible posit arithmetic by reusing existing instructions and implementing custom ISA extensions.}
    \rRevLQ{Experiments show that our design archives up to 2.54$\times$ throughput improvement for posit operations in various GEMM kernels than prior work \cite{cococcioni2021lightweight}.}
\end{enumerate}
\section{Background}\label{sec:bkg}

\begin{figure}[ht]
    \centering
    \includegraphics[width=\linewidth]{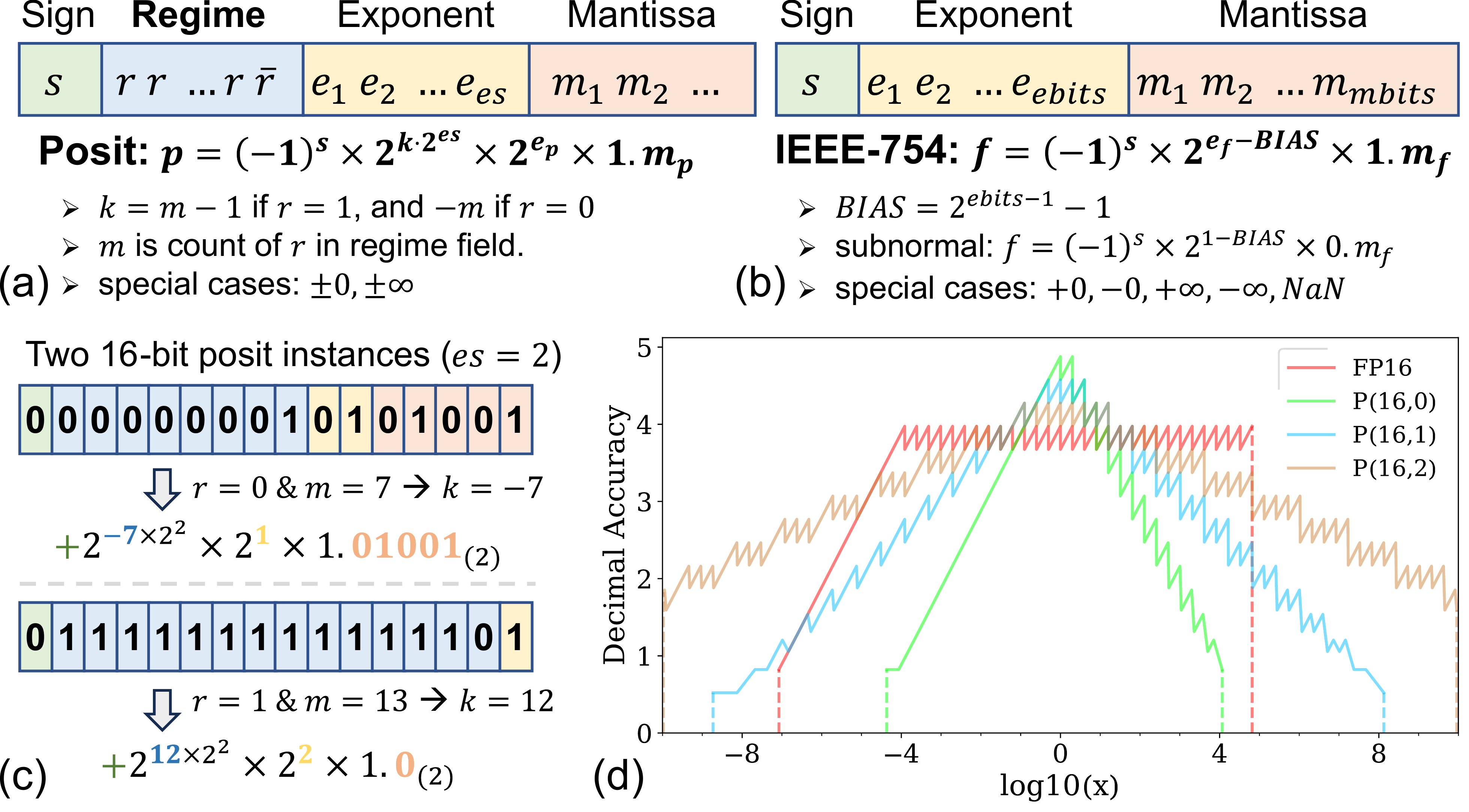}
    \caption{Comparison between posit and IEEE-754 format. (a) and (b) show the respective decoding manner; (c) provides two P(16,2) decoding instances; (d) shows the decimal accuracy of the two formats under various ranges.}
    \label{fig:posit_format}
    \vspace{-0.2cm}
\end{figure}

Posit format \cite{gustafson2017beating}, denoted as P($n$,$es$), is defined by precision ($n$) and exponent size ($es$), comprising four fields: sign, regime, exponent, and mantissa.
It is distinguished from conventional IEEE-754 formats by the unique regime field, which consists of consecutive identical bits followed by an opposite bit, encoding an exponent scale factor.
\rTwoFC{The codec manner of posit and conventional IEEE-754 formats is illustrated in Fig. \ref{fig:posit_format}(a) and (b), respectively.}
Fig. \ref{fig:posit_format}(c) showcases two P(16,2) decoding instances.
Fig. \ref{fig:posit_format}(d) highlights a key advantage of posits, namely their symmetrical tapered accuracy, which can be further adjusted by varying $es$ values.
This flexibility allows for tailored trade-offs between accuracy and dynamic range based on specific application requirements, making posit a compelling alternative to traditional floating-point formats.

Consequently, a growing body of research has emerged around this novel format, including deep learning \cite{lu2020evaluations, ho2021posit, wang2022pl, yu20248, lu2019training}, graphs processing \cite{shah2021dpu}, compute-in-memory \cite{wang202434}, scientific computing \cite{mallasen2023big, klower2019posits}, etc.
\rThreeLQ{Especially, PAUs have drawn increasing attention,}
with studies exploring efficient implementations for basic operations \cite{zhang2020design, zhang2019efficient, chaurasiya2018parameterized, jaiswal2019pacogen, crespo2022unified}, dot-product \cite{li2023pdpu}, 
and approximate designs \cite{gohil2021fixed, kim2024area, zhang2022variable}.
\rThreeLQ{However, they cannot be deployed for real-world computations without system integration.}
\rThreeLQ{Hence, several works \cite{tiwari2021peri, arunkumar2020perc, oh2023rf2p, mallasen2022percival, ciocirlan2021accuracy, rossi2023fppu, sharma2023clarinet, cococcioni2021lightweight} integrate posit arithmetic into RISC-V cores, yet they fall short in hardware efficiency, transprecision capability, and IEEE-754 compatibility.}

\section{The Proposed Architecture}\label{sec:design}
\subsection{\rRevLQ{Lightweight Posit-Enabled RISC-V Core}}

\begin{figure}[t]
    \centering
    \includegraphics[width=\linewidth]{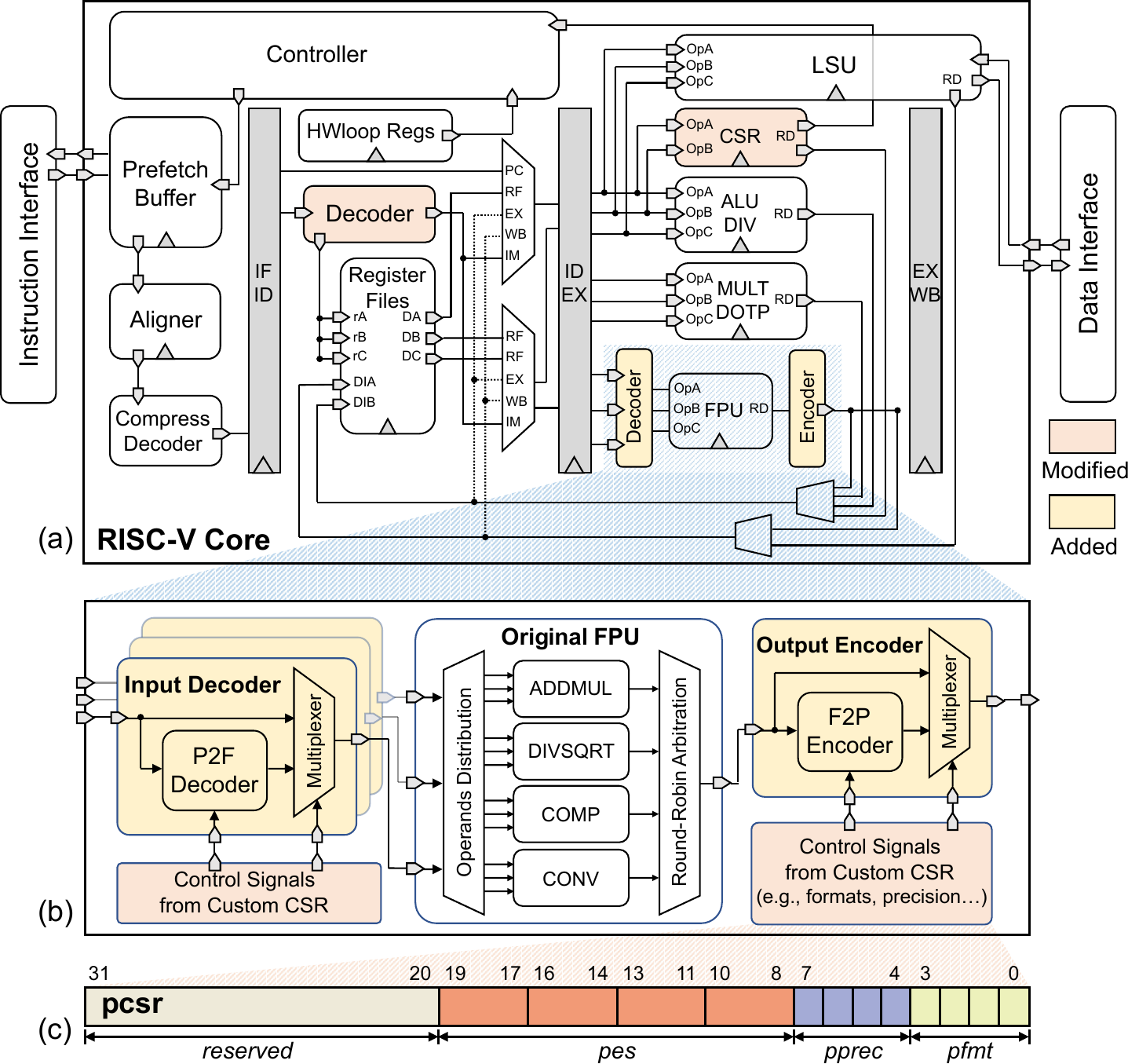}
    \caption{(a) Architecture of posit-enabled RI5CY RISC-V core, with a zoom on (b) extended FPU that supports unified posit/IEEE-754 arithmetic and (c) custom posit control and status register (\textit{pcsr}).}
    \label{fig:riscv_core}
    \vspace{-0.5cm}
\end{figure}

\rTwoFC{Fig.~\ref{fig:riscv_core}(a) illustrates the architecture of our proposed posit-enabled RISC-V core, based on RI5CY~\cite{gautschi2017near}}, featuring an embedded FPU~\cite{mach2020fpnew} that supports IEEE-754 mandated floating-point operations.
\rTwoFC{To seamlessly integrate posit arithmetic, we incorporate dedicated codecs at the FPU's I/O interface, as shown in Fig.~\ref{fig:riscv_core}(b).}
\rTwoFC{The input decoder and output encoder perform posit-to-FP (P2F) and FP-to-posit (F2P) conversions, respectively.}
\rTwoFC{The codecs can be entirely bypassed or selectively disabled, ensuring compatibility with IEEE-754 FP formats and enabling computations between different formats.}
Compared to approaches that replace the original FPU with PAU \cite{rossi2023fppu, arunkumar2020perc, oh2023rf2p, tiwari2021peri, ciocirlan2021accuracy} or embed them in parallel within the RISC-V core pipelines~\cite{mallasen2022percival, cococcioni2021lightweight, sharma2023clarinet}, 
\rThreeLQ{our design achieves unified support for posit and IEEE-754 arithmetic,}
providing complete functionality while minimizing hardware overhead.

We further implement a custom posit control and status register (CSR), namely \textit{pcsr}, 
to facilitate dynamic control of posit operations during program execution.
As depicted in Fig. \ref{fig:riscv_core}(c), \textit{pcsr} holds a 4$\times$1-bit \textit{pfmt} field, a 4$\times$1-bit \textit{pprec} field, and a 4$\times$3-bit \textit{pes} filed, 
indicating the runtime configurations for three input operands and one output result, respectively.
Specifically, \textit{pfmt} determines the currently active format between posit and FP,
essentially controlling whether to \rThreeLQ{skip} the posit codecs during operations.
\textit{pprec} and \textit{pes} allow for adjustments to precision and exponent size for posits, respectively,
which \rTwoLQ{are} well supported by our enhanced microarchitecture.
\rRevLQ{Additionally, we utilize the existent FP CSRs to save exception flags and rounding modes, and reuse FP register files for posit operands, to avoid additional resources.}


\subsection{\rRevLQ{Efficient Precision-Scalable Posit Computing}}\label{sec:micro_archi}
\begin{figure}[t]
    \centering
    \includegraphics[width=\linewidth]{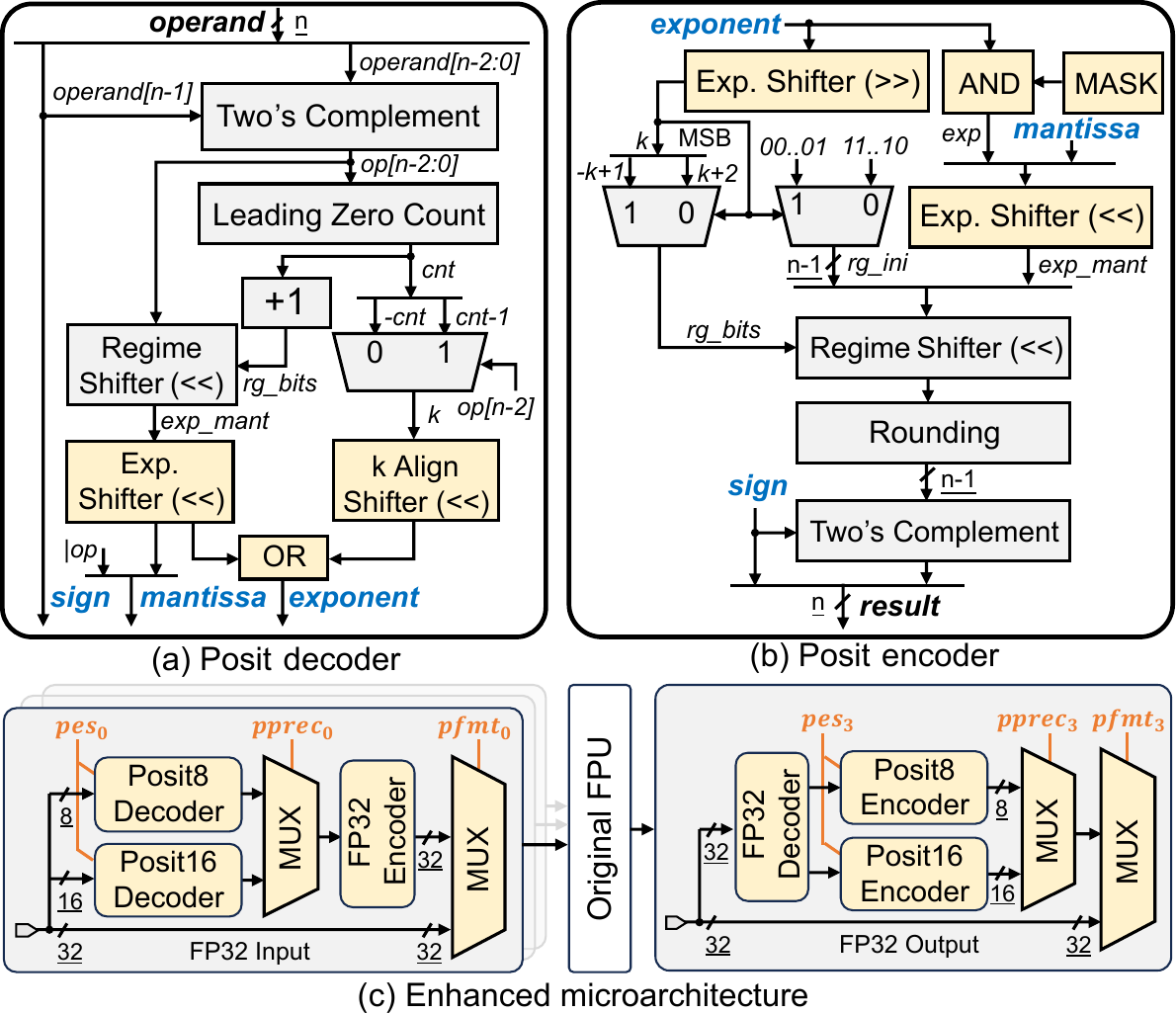}
    \caption{\rRevLQ{Enhanced microarchitecture that supports dynamic exponent size and multi/mixed-precision posit/IEEE-754 arithmetic.}}
    \label{fig:microarchitecture}
    \vspace{-0.4cm}
\end{figure}

\textbf{Supporting \rTwoLQ{dynamic} exponent size.}
\rThreeLQ{Our codecs support this feature to suit different data distribution characteristics.}
Specifically, the posit decoder extracts the valid sign, exponent, and mantissa from the input data.
\rThreeLQ{For fixed $es$, the exponent and mantissa can be derived directly once \textit{exp\_mant} and \textit{k} in Fig. \ref{fig:microarchitecture}(a) are computed.}
\rThreeLQ{However, dynamic left-shifting logic determined by $es$ is introduced to accommodate arbitrary $es$ value for posits, as shown in Fig. \ref{fig:microarchitecture}(a).}
Conversely, the encoder packs the valid components into posit result.
\rThreeLQ{To handle dynamic $es$,}
the encoder calculates effective \textit{exp} and \textit{k} \rThreeLQ{values} using masking and shifting operations.
An additional shifter \rThreeLQ{then} removes invalid bits from the exponent field
\rThreeLQ{before shifting the regime field,}
\rThreeLQ{as shown in Fig. \ref{fig:microarchitecture}(b).}
\rThreeLQ{These hardware modifications allow our design to achieve an optimal balance between accuracy and range for posit formats, thus improving computational accuracy for transprecision computing tasks.}

\textbf{\rThreeLQ{Supporting multi/mixed-precision arithmetic.}}
\rRevLQ{Our design enables 8- and 16-bit posit operations by instantiating the respective codecs in parallel, with \textit{pprec} and \textit{pfmt} controlling their multiplexing, as depicted in Fig. \ref{fig:microarchitecture}(c).}
\rRevLQ{While hardware segmentation techniques \cite{tan2023low} could be used to achieve a unified codec for various precisions, this approach is not implemented since extra control logic negates the benefits of hardware reuse and introduces latency.}
\rThreeLQ{Notably, we intentionally exclude 32-bit posit support to avoid excessive overhead and potential numerical overflow during conversions to FP32.}
\rThreeLQ{Additionally, our design inherently supports mixed-precision and inter-format operations through the \textit{pcsr} configuration, showcasing its excellent transprecision computing capabilities.}


\subsection{\rRevLQ{Instructions for IEEE-754 Compatible Posit Arithmetic}}
\textbf{Reuse instructions for posit arithmetic.}
\rTwoLQ{We reuse the RISC-V standard CSR and ``F" extension instructions to perform posit operations.}
Specifically, \rTwoLQ{FP computations utilize the FP32 format by default.}
However, modifying the \textit{pfmt} field in \textit{pcsr} \rTwoLQ{using CSR write instruction}
switches the computations to posit,
\rTwoLQ{where the precision and exponent size can also be adjusted by updating \textit{pprec} and \textit{pes} fields, respectively.}
This approach avoids additional encoding spaces compared to introducing entirely new instruction sets \cite{mallasen2022percival}.

\textbf{Customize instructions for format conversion.}
\rTwoLQ{We also customize} three types of RISC-V ``F" extension instructions utilizing the available \textit{funct5} field,
\rTwoLQ{to facilitate conversion between different data types and precisions.}
These instructions, detailed in Table \ref{tab:instruction}, are \texttt{fcvt.pfmt.fmt} for converting FP32 to posits, \texttt{fcvt.fmt.pfmt} for converting posits to FP32, and \texttt{fcvt.pfmt.pfmt} for performing mutual conversion between posit formats.
Depending on the decoding of the $es$ field, these operations use either dynamic $es$ value held in \textit{pcsr}, or static $es$ value encoded within the instruction itself.

\textbf{\rThreeLQ{Integrate decoding and compilation support.}}
\rThreeLQ{We further implement decoding and compilation support for custom format conversion instructions, while others leverage existing mechanisms.}
\rThreeLQ{Specifically,} the decoder in Fig. \ref{fig:riscv_core}(a) is enhanced to recognize these new instructions, generating control signals that specify source and destination formats, precisions, and valid $es$ values.
An additional signal is also produced to bypass the unused original FPU to reduce the toggle rate and save power.
\rThreeLQ{Furthermore,}
we modify the GNU \rThreeLQ{Binutils} so that the compiler can recognize the added assembler pseudo instructions.
Therefore, they can be called and compiled alongside the source code by using inline assembly within \rTwoLQ{C programs}.


\begin{table}[t]
\caption{\rOneLQ{Custom Instructions for Conversion between Various Formats and Precisions, with Support for Variable Exponent Size.}}
\label{tab:instruction}
\centering
\resizebox{\columnwidth}{!}{%
\begin{tabular}{@{}lclclclcccc@{}}
\textbf{fcvt.pfmt.fmt} &
\multicolumn{1}{l}{\instbit{31}} &
\multicolumn{1}{r}{\instbit{27}} &
\multicolumn{1}{r}{\hspace{0.10em}\instbit{26}} &
\multicolumn{1}{l}{\hspace{-1.25em}\instbit{25}} &
\multicolumn{1}{r}{\instbit{24}} &
\multicolumn{1}{l}{\instbit{20}} &
\instbitrange{19}{15} &
\instbitrange{14}{12} &
\instbitrange{11}{7} &
\instbitrange{6\hspace{1.3em}}{0} \\ \toprule
fcvt.p8.s    & \multicolumn{2}{c}{0x10} & \multicolumn{2}{c}{0x0} & \multicolumn{2}{c}{0x00} & rs1 & es & rd           & 0x53 \\
fcvt.p16.s   & \multicolumn{2}{c}{0x10} & \multicolumn{2}{c}{0x0} & \multicolumn{2}{c}{0x08} & rs1 & es & rd           & 0x53 \\

\addlinespace[0.8ex]  

\textbf{fcvt.fmt.pfmt} &
\multicolumn{1}{l}{\instbit{31}} &
\multicolumn{1}{r}{\instbit{27}} &
\multicolumn{1}{r}{\hspace{0.10em}\instbit{26}} &
\multicolumn{1}{l}{\hspace{-1.25em}\instbit{25}} &
\multicolumn{1}{r}{\instbit{24}} &
\multicolumn{1}{l}{\instbit{20}} &
\instbitrange{19}{15} &
\instbitrange{14}{12} &
\instbitrange{11}{7} &
\instbitrange{6}{0} \\ \midrule

fcvt.s.p8    & \multicolumn{2}{c}{0x12} & \multicolumn{2}{c}{0x0} & \multicolumn{2}{c}{0x00} & rs1 & es & rd           & 0x53 \\
fcvt.s.p16   & \multicolumn{2}{c}{0x12} & \multicolumn{2}{c}{0x0} & \multicolumn{2}{c}{0x08} & rs1 & es & rd           & 0x53 \\

\addlinespace[0.8ex]  

\textbf{fcvt.pfmt.pfmt} &
\multicolumn{1}{l}{\instbit{31}} &
\multicolumn{1}{r}{\instbit{27}} &
\multicolumn{1}{r}{\hspace{0.10em}\instbit{26}} &
\multicolumn{1}{l}{\hspace{-1.25em}\instbit{25}} &
\multicolumn{1}{r}{\instbit{24}} &
\multicolumn{1}{l}{\instbit{20}} &
\instbitrange{19}{15} &
\instbitrange{14}{12} &
\instbitrange{11}{7} &
\instbitrange{6}{0} \\ \midrule

fcvt.p8.p8   & \multicolumn{2}{c}{0x11} & \multicolumn{2}{c}{0x0} & \multicolumn{2}{c}{0x00}   & rs1 & es & rd           & 0x53 \\
fcvt.p8.p16  & \multicolumn{2}{c}{0x11} & \multicolumn{2}{c}{0x0} & \multicolumn{2}{c}{0x08}   & rs1 & es & rd           & 0x53 \\


fcvt.p16.p8   & \multicolumn{2}{c}{0x11} & \multicolumn{2}{c}{0x1} & \multicolumn{2}{c}{0x00}   & rs1 & es & rd           & 0x53 \\
fcvt.p16.p16  & \multicolumn{2}{c}{0x11} & \multicolumn{2}{c}{0x1} & \multicolumn{2}{c}{0x08}   & rs1 & es & rd           & 0x53 \\

\bottomrule
\end{tabular}%
}
\vspace{-0.4cm}
\end{table}

\section{Experimental Results}\label{sec:res}
To evaluate the hardware impact of our architectural modifications,
we synthesize the individual posit-enabled FPU and RISC-V core under typical operating conditions (0.9 V, \SI{25}{\degreeCelsius}) using Synopsys Design Compiler with the TSMC 28~nm process.
Implemented results on Xilinx ZCU102 FPGA using Vivado 2018.3 are also provided.
Additionally, we assess the system-level impact by analyzing the performance of PULP SoC~\cite{pullini2019mr} \rOneLQ{that incorporates} our modified RISC-V core.
\rTwoFC{Finally, a comprehensive comparison with state-of-the-art (SOTA) posit-enabled RISC-V processors is conducted to highlight the effectiveness of our design.}

\subsection{\rRevLQ{FPU-level Evaluation}}
\rRevLQ{FPUs with varying posit capabilities are evaluated, including original FPU (baseline), +P8 (support 8-bit posit), +MP (support multi-precision posits), and +ES (support dynamic $es$).}
\rThreeLQ{Note that FPU could be internally pipelined to achieve a higher frequency. By default, the embedded ADDMUL, DIVSQRT, COMP, and CONV operation blocks in Fig. \ref{fig:riscv_core}(b) are equipped with 2, 1, 0, and 1 levels of pipeline registers, respectively.}

\rThreeLQ{Table \ref{tab:fpu_dc_synthesis} presents the ASIC synthesis results.
It shows that integrating basic posit support (+P8) only increases 6.3\% timing and 16.4\% area under the default pipeline scheme.
Adding multi-precision and variable $es$ capabilities requires additional resources,
which are modest compared to the original FPU's footprint.
Specifically, latency and area increase by 23.4\% and 16.7\% respectively in implementation with all features (FPU\,+MP\,+ES).}
\rThreeLQ{In contrast, integrating a complete 8- and 16-bit PAU would increase the area by more than 50\% \cite{zhang2019efficient}.}
\rThreeLQ{Moreover, our posit-enabled FPU achieves the same frequency as the baseline with a moderate 19.6\% area overhead, due to strategically inserting an additional pipeline stage.}

\begin{figure}[t]
    \centering
    \includegraphics[width=\linewidth]{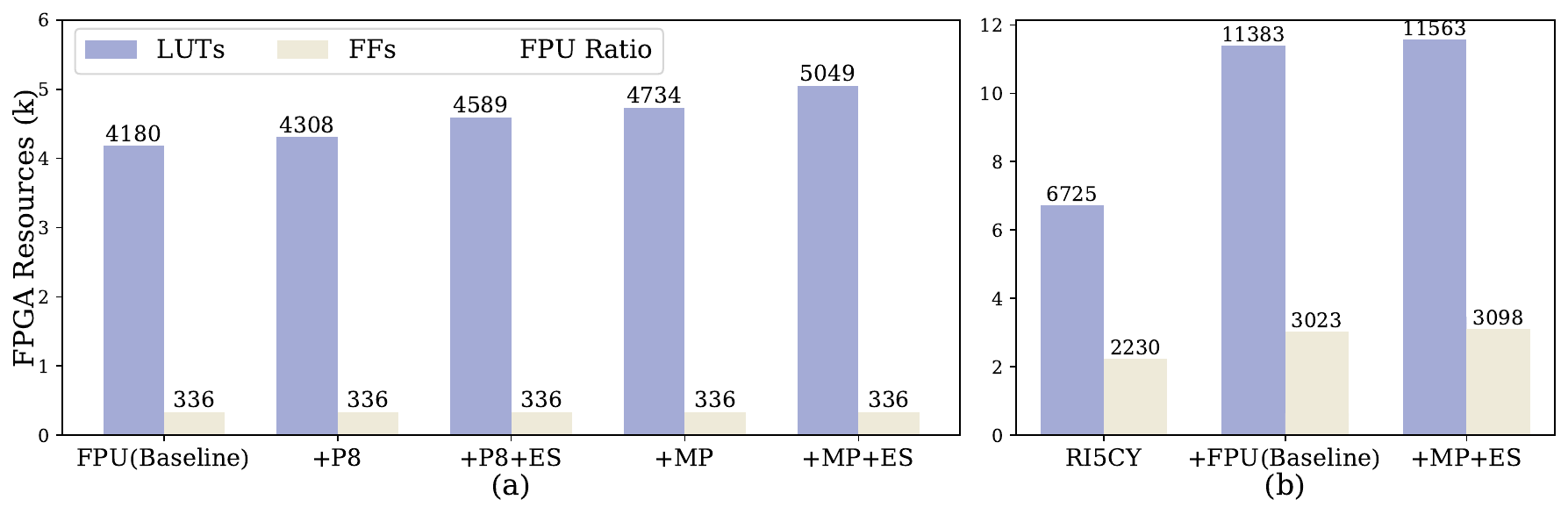}
    \caption{FPGA implemented results for modified (a) FPU and (b) RI5CY core.}
    \label{fig:fpga_resources}
    \vspace{-0.4cm}
\end{figure}

\rThreeLQ{Fig. \ref{fig:fpga_resources}(a) further summarizes the FPGA implemented results of the enhanced FPU with a 6 ns timing constraint.
It shows that enabling P8 support and full functionality increases LUTs at the FPU level by 3.1\% and 20.8\% respectively, 
demonstrating the lightweight nature of our modifications.}

\begin{table*}[th]
\centering
\resizebox{0.95\textwidth}{!}{%
\begin{threeparttable}
\caption{Comparison of the Proposed Design with the State-of-the-art \rTwoFC{Posit-enabled RISC-V Processors}}
\label{tab_comparison}
\begin{tabular}{cccccccc}
\toprule
Implementations & PERCIVAL \cite{mallasen2022percival} & PPU-light \cite{cococcioni2021lightweight} & CLARINET \cite{sharma2023clarinet} & FPPU \cite{rossi2023fppu} & PERC \cite{arunkumar2020perc} & POSAR \cite{ciocirlan2021accuracy} & \textbf{This work} \\
\midrule
RISC-V Core & CVA6 & Ariane & Flute & Ibex & Rocket & Rocket & \textbf{RI5CY} \\
PAU Integration\tnote{$\dagger$} & Parallel & Parallel & Parallel & Replaced & Replaced\textsuperscript{*} & Replaced & \textbf{Unified} \\
Functionality\tnote{$\ddagger$} & ``F" arithmetic & Format Trans. & Fused, Trans. & ``F" arithmetic & ``F" arithmetic & ``F" arithmetic & \textbf{``F" arithmetic} \\
Prec. (Posit \textbar{} FP) & 32 \textbar{} 32 & 8, 16 \textbar{} 32 & 8 (16, 32) \textbar{} 32 & 8 (16) \textbar{} × & 32, 64 \textbar{} × & 32 \textbar{} × & \textbf{8, 16 \textbar{} 32} \\
Transprecision\tnote{$\S$} & × \textbar{} × \textbar{} × & \checkmark \textbar{} × \textbar{} × & \checkmark \textbar{} × \textbar{} × & × \textbar{} × \textbar{} × & \checkmark \textbar{} × \textbar{} × & × \textbar{} × \textbar{} × & \checkmark \textbar{} \textbf{\checkmark \textbar{} \checkmark} \\
\midrule
\begin{tabular}[c]{@{}c@{}}Overhead\\(FPU level)\end{tabular}  & \begin{tabular}[c]{@{}c@{}}+132\% LUTs\\+135\% FFs\end{tabular} & N/A & N/A & × & +26\% LUTs & N/A & \textbf{\rThreeLQ{+20.8\% LUTs}} \\
\midrule
\begin{tabular}[c]{@{}c@{}}Overhead\\(Core level)\end{tabular} & \begin{tabular}[c]{@{}c@{}}+23.9\% LUTs\\+11.3\% FFs\end{tabular} & +1.2\% LUTs & \begin{tabular}[c]{@{}c@{}}+8.6\% LUTs (8b)\\+25.7\% LUTs (16b)\\+71.4\% LUTs (32b)\end{tabular} & \begin{tabular}[c]{@{}c@{}}LUTs\\+7.5\% (8b)\\+17.7\% (16b)\end{tabular} & N/A & \begin{tabular}[c]{@{}c@{}}+30\% LUTs\\-12\% FFs\\+27\% DSPs\end{tabular} & \textbf{\begin{tabular}[c]{@{}c@{}}\rThreeLQ{+1.6\% LUTs}\\+2.5\% FFs\end{tabular}}  \\
\midrule
\begin{tabular}[c]{@{}c@{}}\rThreeLQ{Speedup}\\(\rThreeLQ{SoC level})\end{tabular} & \begin{tabular}[c]{@{}c@{}}GEMM\\(16$\times$16)\\0.77$\times$\end{tabular} & \begin{tabular}[c]{@{}c@{}}DNN Infer.\\0.39$\times$ (8b)\\0.18$\times$ (16b)\end{tabular} & \begin{tabular}[c]{@{}c@{}}GEMM (16$\times$16)\\1.76$\times$ (8b)\\1.72$\times$ (16b)\\1.54$\times$ (32b)\tnote{$\P$}\end{tabular} & N/A & N/A & \begin{tabular}[c]{@{}c@{}}GEMM\\(182$\times$182)\\$\approx$1.0$\times$\end{tabular} & \begin{tabular}[c]{@{}c@{}}\textbf{\rRevLQ{GEMM, GEMV}}\\ \textbf{\rRevLQ{(4$\times$4$\sim$32$\times$32)}}\\ \textbf{\rRevLQ{Softmax (8$\sim$128)}}\\ \textbf{\rRevLQ{$\approx$1.0$\times$}}\end{tabular} \\
\bottomrule
\end{tabular}
\begin{tablenotes}
    \item[$\dagger$] Here ``parallel" and ``replaced" represent that PAU is embedded in parallel with the original FPU in RISC-V cores and that PAU replaces FPU, respectively, while ``unified" means that posit and FP formats are supported in a unified unit. The mark * implies that instruction support is not implemented.
    \item[$\ddagger$] \rThreeLQ{``F" arithmetic means supporting complete functionality, while ``Trans" and ``Fused" imply supporting format transition and fused operations, respectively.}
    \item[$\S$] Indicating whether supporting multi-precision \textbar{} mixed-precision \textbar{} runtime-configurable exponent size, respectively.
    \item[$\P$] The improved performance benefits from the integrated wide accumulator, which incurs excessive overhead and \rThreeLQ{decreases area efficiency.}
\end{tablenotes}
\end{threeparttable}
}%
\vspace{-0.6cm}
\end{table*}

\begin{table}[t]
\caption{\rOneLQ{ASIC Synthesis Results for Enhanced FPU and RI5CY RISC-V Core with Varying Posit Capabilities.}}
\label{tab:fpu_dc_synthesis}
\centering
\resizebox{\columnwidth}{!}{%
\begin{threeparttable}
\begin{tabular}{ccccc}
\toprule
\multirow{2}{*}{Implementations} & \multicolumn{2}{c}{{[Default Pipeline]}} & \multicolumn{2}{c}{{[Pipeline + 1]\tnote{$\dagger$}}} \\
                   & Delay ($ns$) & Area ($um^2$) & Delay ($ns$) & Area ($um^2$) \\
\midrule
\textbf{FPU\,(Baseline)} & \textbf{0.64} & \textbf{11808} & \textbf{0.64}\tnote{$\ddagger$} & - \\
FPU\,+P8 & 0.68 & 13740 & 0.64 & 12509 \\
\rThreeLQ{FPU\,+MP} & 0.73 & 13847 & 0.64 & 14032 \\
FPU\,+P8\,+ES & 0.73 & 12755 & 0.64 & 12949 \\
\textbf{\rThreeLQ{FPU\,+MP\,+ES}} & \textbf{0.79} & \textbf{13775} & \textbf{0.64} & \textbf{14124} \\
\midrule
RI5CY\,(NO FPU) & 0.61 & 24736 & - & - \\
\textbf{RI5CY\,(Baseline)} & \textbf{0.64} & \textbf{39755} & - & - \\
\textbf{RI5CY\,+MP\,+ES} & 0.86 & 41264 & \textbf{0.65} & \textbf{44055} \\
\bottomrule
\end{tabular}
\begin{tablenotes}
    \item[$\dagger$] \rThreeLQ{Inserting one more pipeline stage into the FPU than the default pipeline scheme.}
    \item[$\ddagger$] Timing does not improve further, since DIVSQRT is an iterative block, where adding registers only alleviates the relatively long pre/post-processing paths.
\end{tablenotes}
\end{threeparttable}
}%
\vspace{-0.2cm}
\end{table}

\subsection{\rRevLQ{Core-level Evaluation}}
\rRevLQ{The modified RI5CY core is synthesized under identical conditions.}
As shown in Table \ref{tab:fpu_dc_synthesis}, the original FPU (including 32$\times$32-bit register files) incurs a significant area overhead of 60.7\% at the core level.
On this basis, our enhancements lead to a modest 10.8\% area increase \rTwoFC{without impacting operating frequency.}
\rTwoFC{For ZCU102 FPGA implementation, the core is with an operating frequency constraint of 125 MHz.}
As shown in Fig. \ref{fig:fpga_resources}(b), \rThreeLQ{the impact at the core level is negligible,}
with only a 1.6\% and 2.5\% increase in LUTs and FFs, respectively.
These findings further substantiate the minimal hardware overhead of our design.

\subsection{\rRevLQ{SoC-level Evaluation}}
\rRevLQ{To evaluate the performance at the SoC level},
we implement various general-matrix-multiplication (GEMM) kernels,
and measure execution cycles \rTwoLQ{using cycle-accurate simulation~\cite{bertaccini2024minifloats}} with QuestaSIM by running these kernels
\rTwoLQ{on our enhanced PULP SoC.}
\rTwoLQ{Note that the system runs at 50 MHz on ZCU102 board,}
and we maintain the default pipeline scheme for the built-in FPU since our modifications do not affect the critical path.
Moreover, we simulate the approach in \cite{cococcioni2021lightweight},
\rTwoLQ{where posit values are converted to FP data before further computations.}


\begin{table}[th]
\centering
\resizebox{0.95\columnwidth}{!}{%
\begin{threeparttable}
\caption{Cycles and Throughput when Running Various GEMM Kernels.}
\label{tab_instr_cycles}
\begin{tabular}{cccccccc}
\toprule
 & Prec. & 4$\times$4 & 8$\times$8  & 12$\times$12 & 16$\times$16 & 20$\times$20 & Throughput\tnote{$\ddagger$} \\
\midrule
\multirow{3}{*}{\textbf{Ours}} 
& Baseline & 227 & 1968 & 6307  & \faTimesCircle\tnote{$\dagger$} & \faTimesCircle\tnote{$\dagger$} & 27.20 \\
& \textbf{P(16,1)} & \textbf{231} & \textbf{1972} & \textbf{6311}  & \textbf{14250} & \faTimesCircle\tnote{$\dagger$}     & \textbf{27.02}      \\
& \textbf{P(8,0)}  & \textbf{235} & \textbf{1879} & \textbf{6274}  & \textbf{14201} & \textbf{66609} & \textbf{27.34}      \\
\cmidrule{1-2}
\multirow{2}{*}{\cite{cococcioni2021lightweight}} & P(16,1) & 676 & 4856 & 15488  & 35688 & \faTimesCircle\tnote{$\dagger$}     & 10.39      \\
 & P(8,0)  & 633 & 4798 & 15069  & 34937 & 92586 & 10.75      \\
\bottomrule
\end{tabular}
\begin{tablenotes}
    \item[$\dagger$] \faTimesCircle~means ``illegal instruction" errors occur due to data address overflow.
    \item[$\ddagger$] \rThreeLQ{Average throughput under 4$\times$4$\sim$12$\times$12 GEMM kernels (unit: MFLOPS).}
\end{tablenotes}
\end{threeparttable}
} %
\vspace{-0.2cm}
\end{table}

As shown in Table \ref{tab_instr_cycles},
\rOneLQ{our modified system maintains comparable performance for posits with \rThreeLQ{FP32} baseline,}
since configuring \textit{pcsr} in advance incurs negligible execution cycles. 
\rThreeLQ{In contrast, \cite{cococcioni2021lightweight} suffers from severe performance degradation due to the overhead of two additional type conversion instructions for each operation.}
\rThreeLQ{When running various GEMM kernels, our design achieves a competitive average throughput of 27.34 MFLOPS for 8-bit posit, outperforming \cite{cococcioni2021lightweight} by 2.54$\times$.}


\rThreeLQ{Furthermore, the reduced-precision posit arithmetic offers considerable memory savings.
As illustrated in Table \ref{tab_instr_cycles}, under the same scratchpad memory size in the PULP SoC, a larger GEMM kernel (20$\times$20) can be deployed directly per operation when using 8-bit posit compared to native FP32 (12$\times$12).}

\rRevLQ{We further evaluate GEMV and softmax benchmarks commonly found in DNNs under identical conditions.}
\rRevLQ{Our design achieves 28.1 MFLOPS average throughput for 8-bit posit GEMV operations, excelling \cite{cococcioni2021lightweight} by 2.46$\times$.}
\rRevLQ{It also demonstrates comparable performance ($\approx$ 1$\times$) to FP32 baseline for posit-based softmax kernels, highlighting its practical benefits.}

\subsection{\rTwoFC{Comparison to SOTA Posit-enabled RISC-V Processors}}
\rRevLQ{We compare our design with SOTA posit-enabled RISC-V processors to further evaluate its effectiveness.}
As shown in Table \ref{tab_comparison}, 
our design stands out due to its excellent transprecision capability from various perspectives.
This is achieved through support for both multi/mixed-precision operations and runtime-configurable exponent size.
Furthermore, this work offers complete posit functionality with dedicated hardware and instruction support.
In contrast, \cite{cococcioni2021lightweight, sharma2023clarinet} are limited to specific format conversions or fused operations, and \cite{arunkumar2020perc} lacks instruction support, 
\rThreeLQ{hindering their application scalability.}

Additionally, 
our design achieves a significant reduction of \rThreeLQ{47.9\% LUTs and 57.4\% FFs} at the FPU level compared with \cite{mallasen2022percival},
\rOneLQ{where both the FPU is FPnew \cite{mach2020fpnew}.}
\rRevLQ{Note that we have excluded additional overhead in \cite{mallasen2022percival} introduced by quire accumulators for a fair comparison.}
This efficiency stems from our innovative unified support for posit/IEEE-754 arithmetic, eliminating the need for a separate PAU alongside the FPU.
While \cite{arunkumar2020perc} has a modest 26\% LUTs increase 
\rThreeLQ{by replacing FPU with PAU,}
our design maintains \rThreeLQ{IEEE-754} compatibility and presents more characteristics for posits. 
Furthermore, at the core level, 
our design incurs negligible overhead (\rThreeLQ{1.6\% LUTs and 2.5\% FFs}) compared to prior works \cite{mallasen2022percival, ciocirlan2021accuracy, sharma2023clarinet}, 
which introduce 23.9\%$\sim$71.4\% additional \rThreeLQ{LUTs} despite using less constrained CVA6 or Rocket cores than our resource-limited RI5CY core \rThreeLQ{targeting IoT endpoint devices}.
\rThreeLQ{These results underscore the superior hardware efficiency of our design.}

\rRevLQ{Table \ref{tab_comparison} also demonstrates that our design achieves comparable performance with the baseline when deploying GEMM, GEMV, and softmax kernels.}
This stands in contrast to existing works \cite{mallasen2022percival, cococcioni2021lightweight} which encounter severe performance degradation of up to 23\% and 82\% respectively.
While \cite{sharma2023clarinet} achieves better performance for posits due to \rTwoLQ{the use} of a wide \rThreeLQ{quire} accumulator for fused operations, 
\rTwoLQ{it} nearly doubles the core's resource requirements, 
sacrificing area efficiency.

\section{Conclusion}\label{sec:concls}
\rRevLQ{This paper presents a lightweight integration of precision-scalable posit arithmetic into RISC-V processors for transprecision computing, while maintaining IEEE-754 compatibility. Our approach features three key innovations: 1) lightweight implementation through dedicated posit codecs integrated into the built-in FPU, 2) efficient multi/mixed-precision support with dynamic exponent size for posit numbers, and 3) customized ISA extensions compatible with IEEE-754 floating-point formats.}
\rRevLQ{Experimental results show that our design reduces 47.9\% LUTs and 57.4\% FFs at the FPU level than existing work \cite{mallasen2022percival}.}
\rRevLQ{It also achieves up to 2.54$\times$ throughput improvement compared to the prior approach \cite{cococcioni2021lightweight} when deploying various GEMM kernels.}


\ifCLASSOPTIONcaptionsoff
  \newpage
\fi



\normalem
\bibliographystyle{IEEEtran}
\bibliography{ref_short}
%




\end{document}